\documentclass[conference]{IEEEtran}
\IEEEoverridecommandlockouts

\usepackage{cite}
\usepackage{multirow}
\usepackage{amsmath,amssymb,amsfonts}
\usepackage{algorithmic}
\usepackage{graphicx}
\usepackage{textcomp}
\usepackage{xcolor}
\usepackage{booktabs}
\usepackage{longtable}
\usepackage{caption}
\usepackage{subcaption}
\usepackage{tikz}
\usepackage{units}
\usetikzlibrary{positioning}

\def\BibTeX{{\rm B\kern-.05em{\sc i\kern-.025em b}\kern-.08em
    T\kern-.1667em\lower.7ex\hbox{E}\kern-.125emX}}

\usepackage{hyperref}
\usepackage{tikz}

\usepackage{xcolor}

\usepackage{lipsum}
\usepackage[pscoord]{eso-pic}
\newcommand{\placetextbox}[3]{
  \setbox0=\hbox{#3}
  \AddToShipoutPictureFG*{
    \put(\LenToUnit{#1\paperwidth},\LenToUnit{#2\paperheight}){\vtop{{\null}\makebox[0pt][c]{#3}}}%
  }%
}%

\begin{document}
\placetextbox{0.5}{1}{This is the author's version of an article that has been published.}
\placetextbox{0.5}{0.985}{Changes were made to this version by the publisher prior to publication.}
\placetextbox{0.5}{0.97}{The final version of record is available at \href{https://doi.org/10.1109/WFCS63373.2025.11077657}{https://doi.org/10.1109/WFCS63373.2025.11077657}}%
\placetextbox{0.5}{0.05}{Copyright (c) 2025 IEEE. Personal use is permitted.}
\placetextbox{0.5}{0.035}{For any other purposes, permission must be obtained from the IEEE by emailing pubs-permissions@ieee.org.}%

\title{On the Prediction of Wi-Fi Performance through Deep Learning\thanks{This work was partially supported by the European Union under the Italian National Recovery and Resilience Plan (NRRP) of NextGenerationEU, partnership on ``Telecommunications of the Future'' (PE00000001 - program ``RESTART'') as well as the Swedish Knowledge Foundation through the research profile NIIT.}}

\author{
\IEEEauthorblockN{Gabriele~Formis\IEEEauthorrefmark{1}\IEEEauthorrefmark{3}, Amanda~Ericson\IEEEauthorrefmark{2}, Stefan~Forsström\IEEEauthorrefmark{2}, Kyi~Thar\IEEEauthorrefmark{2}, Gianluca~Cena\IEEEauthorrefmark{1}, and  Stefano~Scanzio\IEEEauthorrefmark{1}}
\IEEEauthorblockA{\IEEEauthorrefmark{1}\textit{National Research Council of Italy (CNR–IEIIT), Italy}\\
gabrieleformis@cnr.it, gianluca.cena@cnr.it, stefano.scanzio@cnr.it}
\IEEEauthorblockA{\IEEEauthorrefmark{2}\textit{Department of Computer and Electrical Engineering, Mid Sweden University, Sundsvall, Sweden}\\
amanda.ericson@miun.se, stefan.forsstrom@miun.se, kyi.thar@miun.se}
\IEEEauthorblockA{\IEEEauthorrefmark{3}\textit{Politecnico di Torino, Italy}}
}

\maketitle

\begin{abstract}

Ensuring reliable and predictable communications is one of the main goals in modern industrial systems that rely on Wi-Fi networks, especially in scenarios where continuity of operation and low latency are required. 
In these contexts, the ability to predict changes in wireless channel quality can enable adaptive strategies and significantly improve system robustness. 
This contribution focuses on the prediction of the Frame Delivery Ratio (FDR), a key metric that represents the percentage of successful transmissions, starting from time sequences of binary outcomes (success/failure) collected in a real scenario. 

The analysis focuses on two models of deep learning: a Convolutional Neural Network (CNN) and a Long Short-Term Memory network (LSTM), both selected for their ability to predict the outcome of time sequences. 
Models are compared in terms of prediction accuracy and computational complexity, with the aim of evaluating their applicability to systems with limited resources.
Preliminary results show that both models are able to predict the evolution of the FDR with good accuracy, even from minimal information (a single binary sequence). 
In particular, CNN shows a significantly lower inference latency, with a marginal loss in accuracy compared to LSTM.
\end{abstract}


\section{Introduction}

In recent years, 
wireless communication technologies and especially the IEEE 802.11 standard (Wi-Fi) 
\cite{10372393} are gaining more and more ground in industry. 
In fact, compared to wired solutions they permit consistent savings concerning installation costs for cables.
Another reason is that not all applications are compatible with wiring, especially those that require mobility.
A relevant example are Autonomous Mobile Robots (AMR), which are being increasingly used in industrial plants: so that they can operate without restrictions, they need to be integrated in the plant through wireless connections \cite{10538029}.

The physical nature of signals, which rely on radio waves propagating on a shared medium, brings several limitations to wireless technologies. 
They may be subject to interference due to other devices operating within the factory \cite{2021103388} or be blocked by obstacles such as walls and machinery \cite{10317890}. 
By making receivers unable to correctly decode the transmitted signal, these phenomena may lead to unreliable communication.

In addition to the ability to support real-time communications,
industrial networks 
must be reliable, and this holds also for \mbox{Wi-Fi} when used in such scenarios. 
Recent studies have tried to face this problem, by using both statistical methods and machine learning (ML) models to predict the quality of Wi-Fi connections in the immediate future \cite{8580452}. 
Doing so 
can help optimize the Media Access Control (MAC) protocol \cite{9134392}. 
Moreover, in systems where data and commands must arrive to destination timely and reliably (e.g., collaborative robots and AMRs), the ability to detect in advance any possible 
deterioration of the communication quality may reduce downtime and operational delays \cite{8666641}.

In these contexts, it is essential to develop techniques that can predict when the channel quality is expected to worsen. 
A relevant metric 
that can be used to this purpose 
is the Frame Delivery Ratio (FDR), which is the percentage of frames successfully delivered 
to destination
over a given time period. 
The ability to predict the FDR enables strategies aimed at improving network robustness, such as 
switching the link to a better channel 
or reducing contextual best-effort traffic produced by applications and sent on air \cite{Khan2022}. 
Hence, our contribution in this work is to demonstrate that deep learning methods can reliably predict the FDR in a real Wi-Fi installation and we will determine which of our tested models is most suitable to be used on embedded devices. 

The most popular techniques for predicting FDR are statistical models such as moving averages or self-regressive models \cite{formis2023linear,10144122}. 
These approaches have a very low computational cost, but may be sub-optimal in non-stationary environments with complex dynamics. 
For this reason, ML models have been developed in recent years that are very effective at predicting complex time series \cite{9786784}. 
This work focuses on two such models: Convolutional Neural Network (CNN) and Long Short-Term Memory (LSTM). 
CNNs are known for their computational efficiency and ability to detect local patterns in time data \cite{liu2014temporal}. 
At the same time, LSTMs are designed \cite{kanto2024wireless} to capture long-term dependencies in sequences, at the expense of higher resource consumption \cite{stenhammar2024comparison}. 

In this paper, after describing the dataset we relied on and the preprocessing methodology in Section~\ref{sec-Dataset}, the used models, the training process, and the optimization of hyperparameters are presented in Section~\ref{sec-Model and Result}. 
The obtained results are given in Section~\ref{sec-result}, before the concluding remarks.

\begin{figure*}[t]
\includegraphics[width=.85\linewidth]{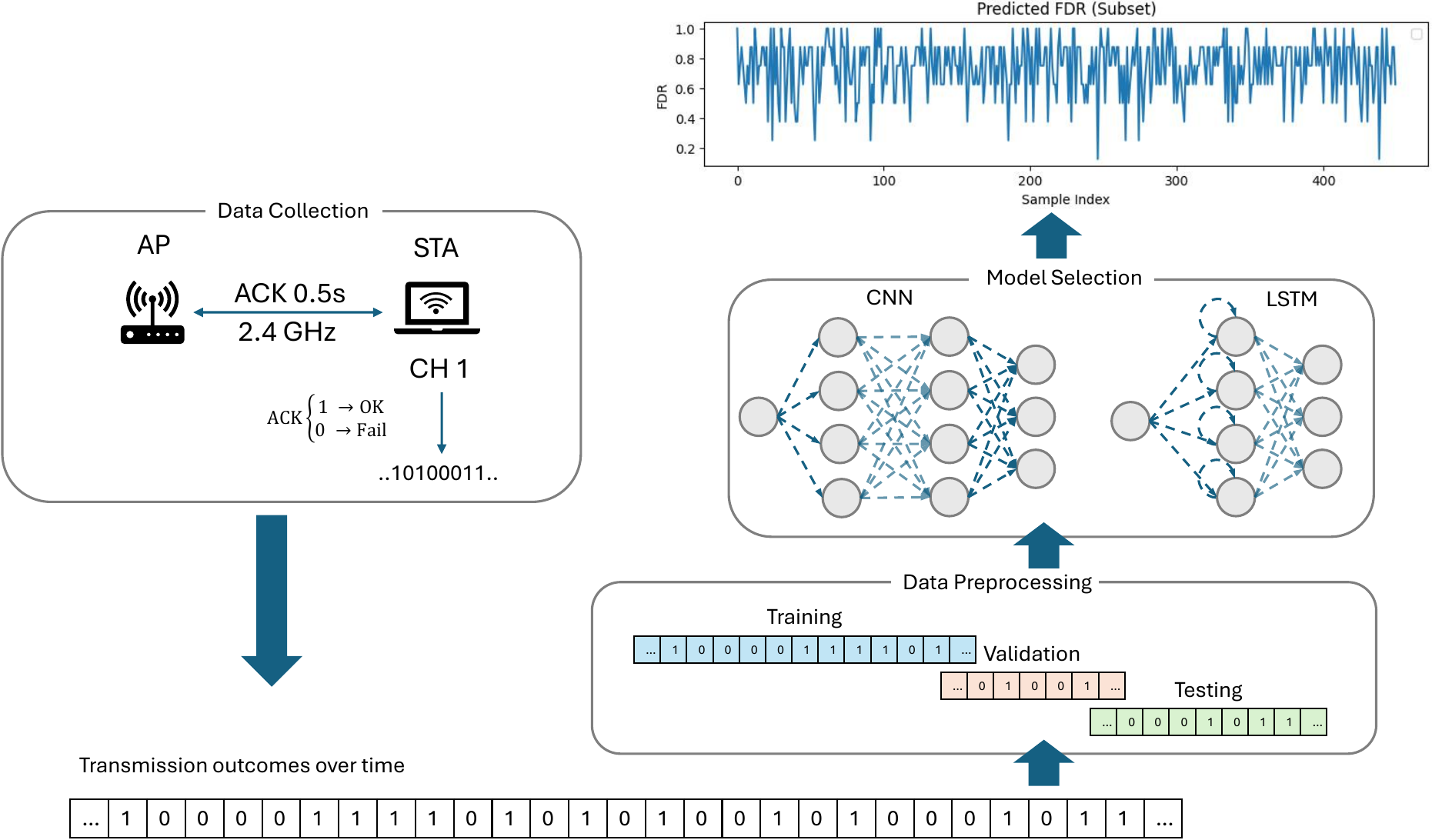}
\centering
\caption{Main steps of the proposed methodology for predicting Wi-Fi network performance.}
\label{fig:pipeline}
\centering
\end{figure*}

\section{Dataset and Preprocessing} \label{sec-Dataset}

The dataset used in this work was collected from a real Wi-Fi communication system. 
However, some features of the IEEE 802.11 protocol have been disabled, including automatic frame retransmission, rate adaptation, backoff mechanisms, and packet aggregation. 
Data is acquired from channel $1$ in the $\unit[2.4]{GHz}$ band. 
The transmission rate has been set to $\unit[54]{Mbps}$, and the system has been configured to repeatedly issue a single transmission attempt every $\unit[0.5]{s}$. 
This allowed the quality of the channel to be sampled at regular intervals. 
Each acquired sample corresponds to a binary outcome: either (\(1\)), if the frame was received correctly and confirmed via ACK, or (\(0\)), if the ACK was not received (transmission failed). 
This binary sequence constitutes our dataset. 
For every discrete time $i$, the target we aim to predict is calculated as
\begin{equation}
    t_i = \frac{1}{N_\mathrm{f}} \sum_{j=i+1}^{i+N_\mathrm{f}} x_j,  
    \label{eq:target}
\end{equation}
where $N_\mathrm{f} = 3600$ ($\unit[30]{min}$) and $x_i \in \{0,1\}$. 
The target $t_i$ represents an estimate of the FDR over the next $\unit[30]{min}$
(examples of its trend in real links are available in \cite{10295470}). 
The dataset was subsequently divided into three parts: training (\mbox{$\sim\!30$} days), validation ($\sim\!10$ days), and test ($\sim\!20$ days).

However, the samples in the dataset showed an imbalance in favor of successful (85\%) versus unsuccessful transmissions (15\%). 
This imbalance makes prediction through ML predictive models more difficult, because they must be able to pick up the weak signals related to failures without falling into a trivial strategy of constant success prediction.
The Synthetic Minority Over-sampling Technique (SMOTE) 
\cite{SMOTE}
has been tested to deal with this problem. This method, used to balance datasets by generating artificial examples of the underrepresented class (0s), helps ML models ``learn'' better by reducing bias toward the majority class. However, it did not provide significant improvements in model accuracy. 
Conversely, it led to a considerable increase in the training time, so we decided to discard it and avoid making changes to the dataset.

In Fig.~\ref{fig:pipeline}, the main steps of our proposed methodology are summarized,
from dataset acquisition to the prediction of the FDR using two different ML models that will be described in detail in the following sections.
In this preliminary work, only the sequence of the binary outcomes $x_i$ has been fed as input to the models. 
The use of other measured quantities, such as the Received Signal Strength Indicator (RSSI) or the transmission latency, will be explored in future works.

\section{Model Architectures and Training Process} 
\label{sec-Model and Result}
Two models were analyzed: CNN and LSTM. 
Both are capable of learning time sequences, but they exhibit different computational and structural characteristics. 
This section analyzes these models, how they were trained, and the choice of hyperparameters. 
Finally, results will be presented.

\begin{table}[t]
\centering
\caption{Model Parameters for CNN and LSTM}
\label{tab:model_parameters_combined}
\small
\tabcolsep=0.14cm
\def\arraystretch{1.0}
\begin{tabular}{|l|c|c|c|c|}
\hline
{Parameter}          & \multicolumn{1}{c|}{{CNN}}       & \multicolumn{1}{c|}{{LSTM}} \\
\hline
{Input Seq. Len.} ($l$) & 3600                       & 1200 \\ 
{Batch Size}  ($b$)   & 64                            & 32      \\ 
{Epochs} ($N_\tau$)   & 30                              & 15      \\ 
{Optimizer}           & Adam                          & Adam     \\ 
{Loss Function}       & MSE  & MSE  \\ 
{Learning Rate}       & 0.01 (decay)          & 0.01 (decay)      \\ 
{Filters/Units ($n$)} & 128 (filters)      & 25 (LSTM units)  \\ 
{Kernel Size}         & 3                               & N/A    \\ 
{Pooling}             & MaxPooling (size 2)  & N/A              \\ 
{Dense Layers}        & 3 (128, 64, 1 unit)  & 1 (1 unit)      \\ 
{Activation Func.} & ReLU (Conv1D/Dense) & Tanh         \\ \hline
\end{tabular}
\end{table}

\subsection{Models Architecture} 

CNNs were initially used for image processing, but have also proven effective in time series analysis. Convolutional filters allow local patterns and repetitive structures to be identified in the data. Another advantage of CNNs is that they speed up training by sharing weights, which reduces the total number of parameters. However, they also have weaknesses, since their ability to capture time relationships decreases as the distance between events increases.

LSTM, on the other hand, is a variant of the Recurrent Neural Networks (RNN). RNNs are designed to process data sequences while maintaining an internal memory that allows them to ``remember'' the past when processing the current inputs. 
For this reason, they are suitable for the prediction of time series where previous information influences the future. 
However, LSTMs have been introduced to address the problem of vanishing gradients, making it difficult to learn long-term dependencies. 
An LSTM consists of three main gates to regulate the flow of information:
the first is the \textit{input gate}, which determines what new information to add to the cell state; 
then, there is the \textit{forget gate}, which decides what information to keep or discard from memory; 
finally, the \textit{output gate} determines which part of the cell state to use as output.
This structure allows LSTMs to store and use information over extended time intervals, making them particularly effective in capturing long-term time dependencies in time series.

\subsection{Training process}

The reference target in Eq.~\eqref{eq:target} represents 
the FDR over the next $30$ minutes, computed as the mean value of outcomes.
What we want is to estimate it from the sequence of past binary observations. 
As a consequence, both models were trained to perform a regression task. 
As the loss function we selected the Mean Squared Error (MSE), a customary choice in regression problems, which penalizes in a quadratic way deviation between the predicted and the actual values. 
The \textit{Adam} optimizer was selected for its efficiency in handling noisy gradients and convergence speed.

During the training phase, the learning rate was decreased from epoch to epoch. 
In particular, it was initially set to $0.01$ and was halved on every subsequent epoch. 
This strategy allows to start the optimization with relatively large updates, and to refine them progressively, improving the stability of training in the final stages.
However, to prevent overfitting phenomena, an early stopping mechanism has been implemented, based on the MSE performance on the validation set. 
In detail, if the loss computed on the validation dataset $V$ does not improve for a certain number of consecutive periods, training is stopped early.

\subsection{Hyperparameter tuning}
Hyperparameter optimization was performed using the \texttt{fmin} function of the \texttt{hyperopt} library, which implements a strategy based on Tree-structured Parzen Estimators (TPE). 
The objective is to minimize a cost function defined as the mean of the MSE on the validation dataset $V$. 
For each model $M$ (CNN or LSTM) and configuration of hyperparameters $\theta$, the objective function is evaluated as
\begin{equation}
J(M, \theta) = \sum_{(p_i, t_i) \in 
 V}\Big( t_i-f_M(p_i,\theta)\Big) ^2,
 \label{eq:loss}
\end{equation}
where $p_i$ represents the input pattern, 
$t_i$ is the target computed through Eq.~\eqref{eq:target} on the validation dataset, 
and $f_M(p_i,\theta)$ is the prediction of the model $M$ with hyperparameters $\theta$.

 \begin{table*}[t]
  \caption{Prediction accuracy of CNN and LSTM models
  (statistics on the prediction error).}
  \label{tab:metrics_lstm}
  \normalsize
  \begin{center}
    \tabcolsep=0.17cm
    \def\arraystretch{1.1}
    \begin{tabular}{c|ccccc|cccccc|cccc}
    Model & $\mu_{e^2}$ & $e^2_{\mathrm{p}_{90}}$ & $e^2_{\mathrm{p}_{95}}$ & $e^2_{\mathrm{p}_{99}}$ & $e^2_{\mathrm{max}}$ & $\mu_{|e|}$ & $\sigma_{|e|}$ & ${|e|}_{\mathrm{p}_{90}}$ & ${|e|}_{\mathrm{p}_{95}}$ & ${|e|}_{\mathrm{p}_{99}}$ & ${|e|}_{\mathrm{max}}$ & ${e}_{\mathrm{min}}$ & ${e}_{\mathrm{p}_{5}}$ & ${e}_{\mathrm{p}_{95}}$ & ${e}_{\mathrm{max}}$ \\
    & \multicolumn{5}{c|}{$[\cdot 10^{-3}]$} & \multicolumn{6}{c|}{[\%]} & \multicolumn{4}{c}{[\%]} \\
    \hline
    CNN & 2.15 & 5.64 & 11.77 & 30.26 & 100.24 & 2.80 & 3.69 & 7.51 & 10.85 & 17.40 & 31.66 & -31.63 & -6.41 & 4.69 & 31.66 \\
    LSTM & 2.08 & 5.59 & 11.68 & 30.00 & 99.90 & 2.77 & 3.67 & 7.48 & 10.82 & 17.35 & 31.50 & -31.60 & -6.39 & 4.67 & 31.50 \\
    \hline
    \end{tabular}
    \end{center}
    \vspace{2mm}
\end{table*}

\begin{table} ]
\caption{Computational complexity of CNN and LSTM } 
\centering
\normalsize
\tabcolsep=0.20cm
\def\arraystretch{1.1}
\begin{tabular}{c|ccc}
 & {Mean} & {Memory} & {Memory} \\
{Model} & {response time} & {footprint} & {peak} \\
& {} ($\unit[]{ms}$) & {} ($\unit[]{MB}$) & {} ($\unit[]{MB}$) \\
\hline
CNN  & 2.7 & 0.02 & 0.03 \\
LSTM  & 20.3 & 0.16 & 0.67 \\
\hline
\end{tabular}
\label{tab:complexity}
\vspace{2mm}
\end{table}

For each epoch  $\tau$, we define $J(M, \theta, \tau)$ as the loss on the validation set, i.e., the MSE between the model prediction and the target values at epoch $\tau$.
This quantity
tends to vary significantly during the early stages of training. For this reason, the average stable loss is calculated only from the sixth epoch,
\begin{equation}
\overline{J}(M, \theta) = \frac{1}{N_\tau-5} \sum_{\tau=6}^{N_\tau}{J(M, \theta, \tau)},
\label{eq:avg_loss}
\end{equation}
where $N_\tau$ is the total number of epochs. This allows different configurations to be compared based on average behavior, excluding the first unstable epochs. 
The selection of optimal hyperparameters $\theta^*$ is then calculated as

\begin{equation}
\theta^* = \arg \min_{\theta} \overline{J}(M, \theta).
\label{eq:opt_single_ch}
\end{equation}

Different hyperparameter values have been explored:
the \textit{batch size} \( b \) was chosen as \( b \in \{32, 64, 128\} \); 
the \textit{number of filters} (for CNN) or \textit{units} (for LSTM), indicated by $n$,
has been selected as $n \in \{64, 128, 256\}$. 
Concerning the \textit{length} $l$ of the input sequence, three values were considered, $l \in \{1200, 1800, 3600\}$, corresponding  to $10$, $15$, and $30$ minutes of observations, respectively. 
The ranges we selected for these parameters reflect a trade-off between model accuracy and convergence stability. 
Higher values were avoided because they often led to unstable training or non-convergence, especially due to increased model complexity and memory demands.
Table~\ref{tab:model_parameters_combined} reports the optimal hyperparameters values we chose for each model, which have been used for evaluating results in the test phase.

\section{Results} \label{sec-result}
This section analyses the performance obtained by the CNN and LSTM models. 
To this aim, we use statistics calculated on the square error $e^2$, absolute error $|e|$, and simple error~$e$. 
Table~\ref{tab:metrics_lstm} provides a comparison between the two models. 
As can be seen, both achieve comparable results.  
LSTM shows a modest advantage in accuracy, with a slightly lower MSE ($\mu_{e^2} = 2.08 \cdot 10^{-3}$ versus $2.15 \cdot 10^{-3}$ from CNN) and a smaller mean absolute error ($\mu_{|e|} = 2.77\%$ versus $2.80\%$).
However, the differences between models are minimal. 
The maximum errors observed are almost identical (about $31.5\%$) and also the values of high percentiles ($|e|_{p95}$ and $|e|_{p99}$) are very similar. 

The average FDR on the channel is relatively high ($88.4\%$), but the proposed methods allow to anticipate its fluctuations, providing a way to adapt the network parameters in real-time and improve reliability in critical instants.
Although LSTM can capture time sequences better than CNN, the benefits in terms of accuracy are negligible.
Conversely, CNN is noticeably more efficient than LSTM in terms of calculation time and memory occupation (see Table~\ref{tab:complexity}). 
Besides being more than seven times faster, 
CNN achieves a significantly lower memory footprint ($\unit[0.02]{MB}$ vs. $\unit[0.16]{MB}$) and memory peak ($\unit[0.03]{MB}$ vs. $\unit[0.67]{MB}$) compared to LSTM. 
Memory footprint refers to the average resident set size (RSS) of the process during inference, while memory peak indicates the maximum memory allocated, including stack and heap, during model execution.

This makes CNN a better choice for the implementation on embedded devices and in real-time systems, while LSTM is preferable when computational resources are not an issue. 
Computational complexity 
was evaluated
on a machine with $\unit[8]{GB}$ RAM and an Intel Core i3-10105 CPU ($\unit[3.7-4.4]{GHz}$).

\section{Conclusions}
Both CNN and LSTM demonstrated good capabilities in predicting the FDR in a real Wi-Fi installation. 
Despite the slightly worse prediction accuracy compared to LSTM,
the CNN method is the best candidate for embedded devices due to its lower computational requirements.

Future work includes devising improved hybrid architectures, such as joint CNN-LSTM models, which can combine advantages in terms of efficiency and time modeling capability. We also plan to explore transfer learning techniques to improve model generalization across different environments, devices, and deployment scenarios.
Besides, additional input features will be considered, such as RSSI and latency, as well as the extension to multiple channels and the validation in the presence of mobility.

\bibliographystyle{IEEEtran}
\bibliography{references}
\end{document}